# Critical dynamics governs deep learning

Simon Vock[1,2,3,4] and Christian Meisel[1,2,4,5] *

[1] Computational Neurology, Department of Neurology, Charité – Universitätsmedizin, Berlin, Germany
[2] Berlin Institute of Health, Berlin, Germany
[3] Faculty of Life Sciences, Humboldt University Berlin, Germany
[4] Bernstein Center for Computational Neuroscience, Berlin, Germany
[5] Center for Stroke Research, Berlin, Germany

*christian@meisel.de

# Abstract

The rapid advances in artificial intelligence (AI) have largely been driven by scaling deep neural networks (DNNs) - increasing model size, data, and computational resources. However, performance is ultimately governed by network dynamics. The lack of a principled understanding of DNN dynamics beyond heuristic-based design has contributed to challenges with their robustness, suboptimal performance, high energy consumption and pathologies in continual and AI-generated content learning. In contrast, the human brain does not seem to suffer these problems, and converging evidence suggest that these benefits are achieved by dynamics being poised at a critical phase transition. Inspired by this principle, we propose that criticality provides a unifying framework linking structure, dynamics, and function also in DNNs. First, by analyzing more than 80 state-of-the-art models, we report that a decade of AI progress has implicitly driven successful networks towards criticality – explaining why certain architectures succeeded while others failed. Second, we demonstrate that incorporating criticality explicitly into training improves robustness and accuracy preventing key limitations of current models. Third, we show that catastrophic AI pathologies, including the performance degradation in continual learning and in model collapse - where performance degrades when training on AI-generated data - constitute a loss of critical dynamics. By maintaining networks at criticality, we provide the first principled solution to this fundamental AI problem, demonstrating how criticality-based optimization mitigates performance degradation and model collapse. This work confirms criticality as substrate-independent principle of intelligence, connecting AI advancement with core principles of brain function. It provides profound theoretical insights along with immediate practical value solving major AI challenges to ensure long-term DNN performance and resilience as models grow in scale and complexity.

# Main text

The progress in artificial intelligence (AI) over the past decade is based on neural networks designed to mimic brain functioning. Much of the improvement in AI capabilities relies on deep neural networks[1] (DNNs) and has been attributed to their scaling: bigger networks with more parameters to train, larger amounts of data, and more computational resources[2,3]. The intuition behind this is simple: greater AI systems, trained with sufficiently large data and computational resources, tend to better represent the information contained in the data, leading to better performance. While size and structure are crucial, the functioning of DNNs and networks in general is ultimately determined by its dynamics: performance will only be optimal when input-driven activity propagates appropriately through the network to elicit the right outputs. Algorithmic improvements shaping DNN dynamics, including changes in model architectures[4,5], optimization algorithms, neuron dynamics and transfer functions[6–10], weight initialization[11–13], software frameworks, have consequently also contributed to AI performance[14]. However, the role of DNN dynamics and how to quantify and optimize it are not well understood. By consequence, everyday decisions including model assembly for optimal task performance, choice of hyperparameters, weight initialization and securing a stable training process, still largely rely on heuristics, trial and error[15–17]. More importantly, this limited understanding of DNN dynamics is at the core of key challenges in AI, including loss of plasticity—diminished performance in continual learning[18]—and model collapse—performance degradation when training on AI-generated content[19]. Generative AI (GenAI), including large language models (LLMs) like ChatGPT are now a lasting force in online content. Problems such as model collapse remain poorly understood and lack principled solutions. As future models train on AI-generated content, understanding and preventing model collapse will be key to ensuring their continued reliability and progress.

The human brain does not seem to suffer these problems. A growing amount of research indicates that cortical network dynamics is best described by a critical state, poised at the phase transition between chaotic and ceasing neuronal activity. Physics and information theory indicate that numerous information processing functions, including information transmission, integration, storage, dynamic range, and sensitivity to inputs, are optimized simultaneously when networks operate near a critical phase transition point[20,21,22]. Empirical evidence supporting the relevance of criticality in cortical networks has come from experimental observations in animals and humans showing power-law scaling[23] and long-range temporal correlations[24,25] which are hallmarks of criticality. Recent work has demonstrated that these signatures are predictably perturbed when networks are moved away from the critical point[26,27], that proximity to critical dynamics predicts cognitive performance across multiple domains[28], and that homeostatic mechanisms like sleep are required to restore the computationally optimal critical state[29–31]. In contrast, deviations from criticality have been linked to neurological diseases[32,33] and cognitive deficits[28,34]. Thus, there is strong and converging empirical evidence for the relevance of criticality as a fundamental framework linking network structure to dynamics and (optimal) function in cortical networks.

Here we show that criticality provides a fundamental framework linking network structure to dynamics and (optimal) function also in deep learning. Drawing inspiration from biological principles, we first show how a critical phase transition characterizes the balance between quiescence and unhindered activity in DNNs. This critical regime allows for optimal information processing and robust yet flexible network dynamics. Using this framework, our work provides three key insights. First, by systematic investigation of 80 modern deep learning models trained on the ImageNet-1k dataset[35], we demonstrate that over the past decade, DNNs have not only become increasingly accurate in their classifications but have also evolved towards more critical dynamics. This observation highlights how model-specific, often heuristic approaches have implicitly adopted these principles to drive the networks towards the critical regime, resulting in balanced signal propagation and optimal error backtracking. Second, building upon these insights, we derive a training scheme that explicitly incorporates criticality, and thereby improves robustness and performance. Third, we show that the performance loss in continual deep learning and in model collapse is caused by loss of criticality. Maintaining close proximity to criticality, conversely, prevents loss of performance in continual learning and mitigates model collapse. Collectively, this biologically informed perspective unifies a wide range of state-of-the-art DNN techniques under a common theoretical framework. It demonstrates that designing effective and performant artificial neural networks requires considering more than just size, and that facilitating critical dynamics within the network is essential for enhancing computational performance.

# A critical phase transition governs the dynamics in deep neural networks

Activity propagation in many biological and artificial neural networks can be adequately described by a branching process[23,38,39]. In such systems, activity remains small and local when interactions are weak, while overly strong interactions lead to network-wide overactivation. At the critical transition between these states, activity propagates in balanced cascades, avoiding both premature die-out and runaway excitation. To demonstrate the generic relevance of a branching process for DNNs and its relationship to optimal function, we first review a simple fully connected DNN (Figure 1a). Two dynamical extremes can be observed when scaling the weights: (1) small weights promoting die-out dynamics where any input vanishes after a small number of layers, (2) large weights resulting in exponentially growing neuron outputs in each layer (Figure 1a, brown and light green boxes). The branching parameter $\sigma$ and maximum Lyapunov exponent $\lambda_0$ fully characterize the dynamics in these regimes: subcritical ($\lambda_0 < 0$ and $\sigma < 1$) and supercritical ($\lambda_0 > 0$ and $\sigma > 1$; Figure 1b). At a critical coupling strength, the phase transition between these two regimes occurs which is characterized by balanced activity propagation (Figure 1a, dark green box). In the vicinity of this critical state ($\lambda_0 \approx 0$, $\sigma \approx 1$, $\Sigma \approx 0$; Figure 1b, dark green box), the network is able to respond to a maximal range of input stimuli, reflected in peak dynamic range $\Delta$, along with various other functional benefits. As predicted by theory and prior experiments[40], the precise connectivity at which the critical state occurs depends also in DNNs on network size (Figure 1c). These finite size effects are thus important to take into consideration when designing and scaling artificial neural networks.

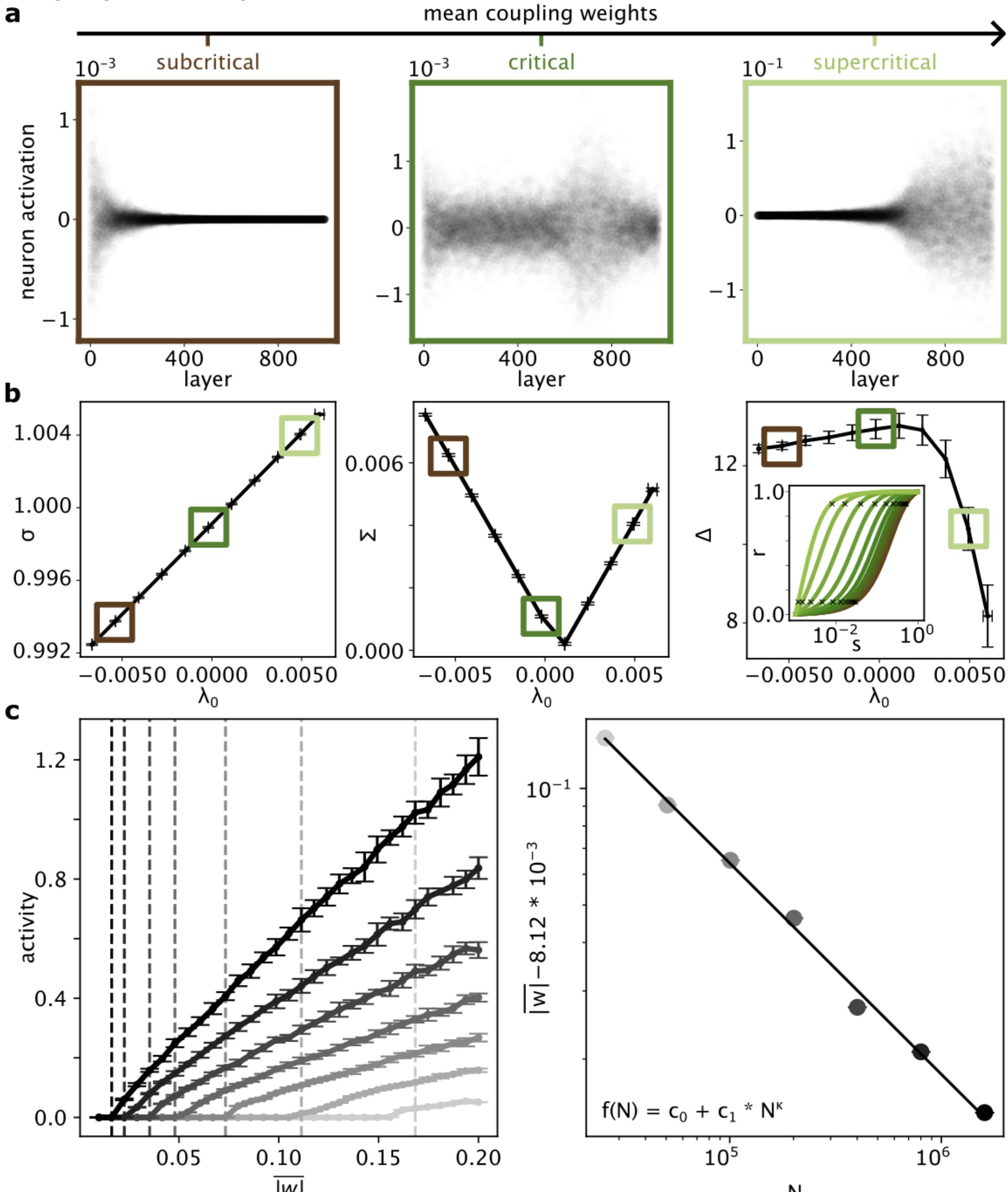

**Figure 1: A critical phase transition governs the dynamics in deep neural networks (DNNs). a**, Neuron activations in DNNs with small (subcritical, brown), medium (critical, dark green), and large (supercritical, light green) mean coupling strengths from a uniform input range $[-10^{-3},\ 10^{-3}]$. **b**, The critical state (dark green) is characterized by balanced activity propagation $\sigma \approx 1, \Sigma \approx 0,$ and a vanishing maximal Lyapunov exponent $\lambda_0 \approx 0$, alongside a maximal dynamic range $\Delta$. Inset: Normalized stimulus-response curves for DNNs with increasing average coupling strengths, brown to green. Black crosses mark stimuli corresponding to 10% and 90% response levels. **c**, Left: Sharply increasing neuron activities at a critical coupling strength $w_c$ for DNNs of increasing size. The critical point is subject to finite size effects, where smaller DNNs (grey) undergo a phase transition at higher coupling strengths than larger DNNs (black). The vertical dashed lines mark the critical coupling strength $w_c$ where $\lambda_{max} = 0$. Right: The shift of $w_c$ to higher

coupling strengths for an increasing number of neurons $N$ is fitted with a power law $f(N) = c_0 + c_1 N^{\kappa}$ where $c_0 = 8.12 \times 10^{-3}$, $c_1 = 3.14 \times 10^{1}$, $\kappa = -5.45 \times 10^{-1}$ (solid line). All DNNs have 1,000 layers and each DNN has either 25, 50, 100, 200, 400, 800, or 1,600 neurons per layer (grey to black). The dynamic range $\Delta$ is calculated from a uniform stimulus range $[-1,\ 1]$ that is scaled logarithmically with $[10^{-3},\ 10^{0}]$.

# DNN evolution over the years: The closer to criticality, the better the performance

We next investigated the relationship between DNN computational performance and proximity to criticality. Computer vision tasks, in particular, have experienced strong performance gains over the last decade on standardized test data, making them a suitable target for this approach. The ImageNet-1K[35] leaderboard represents the state-of-the-art in image classification tasks on the ImageNet dataset. Over the past decade, increasingly refined and specialized network architectures have significantly improved performance on this task since 2012 (Figure 2a; Spearman rank correlation coefficient $r = 0.72$, p-value $3.92 \times 10^{-14}$). We therefore focused on 80 different pre-trained models from the PyTorch Torchvision repository[41], all trained on the ImageNet-1K dataset. These models included various network types and sizes, allowing us to systematically study the relationship between prediction accuracy, DNN dynamics, and structure (see Supplementary Table S1 for model details). With increasing performance we observed a systematic shift towards criticality during the same period, as indicated by decreases in the parameters $\Sigma$ and $\lambda_0$ by more than one order of magnitude (Figure 2a; $\Sigma$: $r = -0.37$, p-value $6.00 \times 10^{-4}$; $\lambda_0$: $r = -0.46$, p-value $1.15 \times 10^{-5}$). This relationship holds across a diverse range of network architectures, including residual networks, convolutional neural networks, vision transformers, and many others.

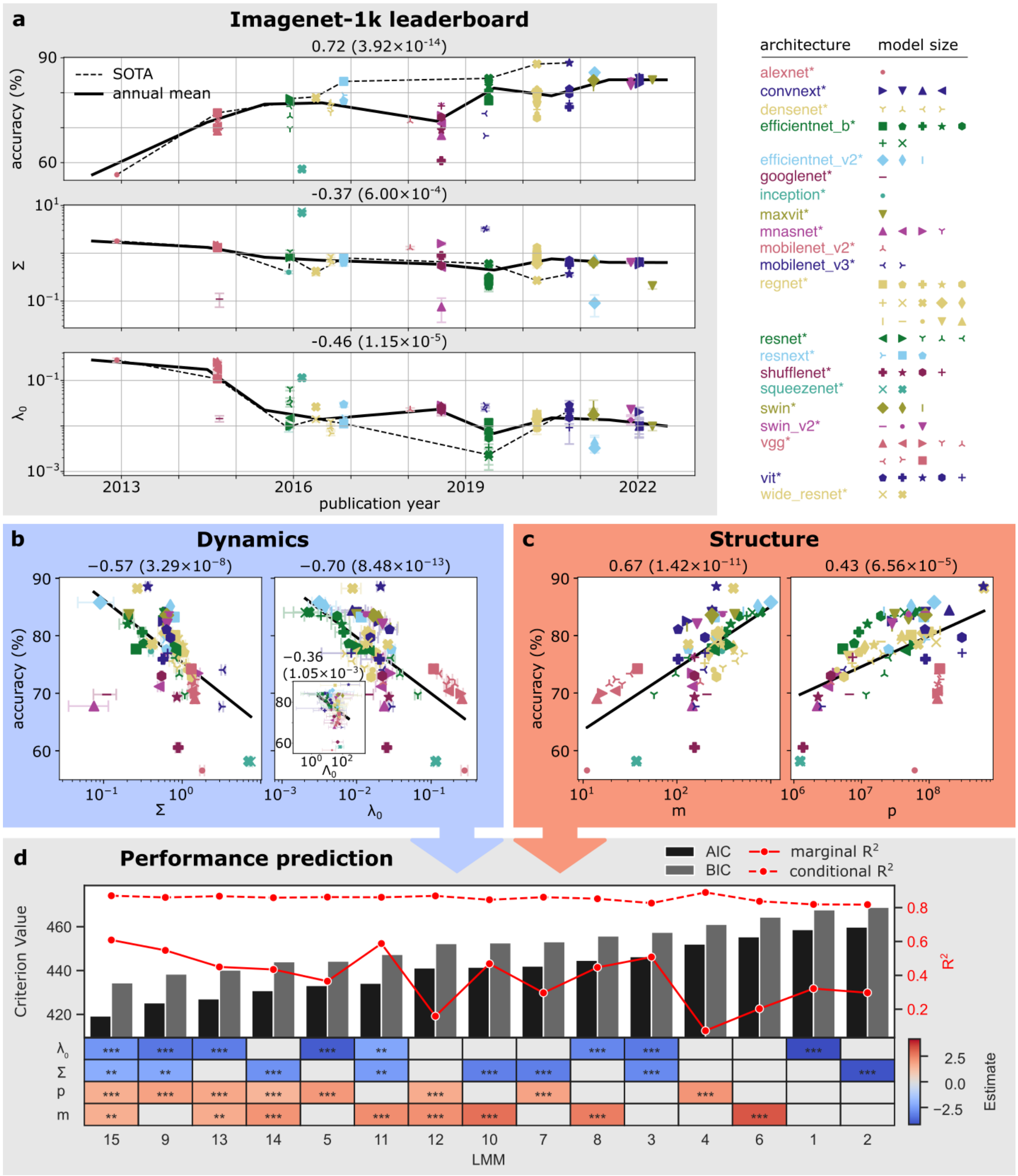

**Figure 2: Historical increases in prediction accuracy are accompanied by more critical dynamics in DNNs. a**, Model performance from 80 DNNs pre-trained on the ImageNet-1k dataset over the last years shows a shift closer to the critical phase transition (midldle and bottom rows) as prediction accuracy increases (top row). Solid lines represent the annual average, while dashed lines denote state-of-the-art (SOTA) networks according to classification accuracy on the validation set. **b**, A strong negative correlation between prediction accuracy and distance to the critical point (measured by $\Sigma$ and $\lambda_0$). Inset: The maximal singular values $\Lambda_0$ are independent of model structure and are significantly negatively correlated with model performance. **c**, Correlation of DNN number of modules $m$ and number of free parameters $p$ with prediction accuracy. **d**, Fit of 15 linear mixed-effects models (LMMs) with $\lambda_0$, $\Sigma$, $p$ and $m$ as fixed effects in all combinations without interaction terms, and DNN architecture as random effects. Colors represent a negative (blue) and positive (red) estimated effect. Statistical significance of correlations is shown with asterisks (*** p < 0.001, ** p < 0.01, * p < 0.05). Marginal (solid red line) and conditional (dashed red line) coefficients of determination $R^2$ denote the proportion of variation that is explained by the fixed effects and the entire model, respectively. The Akaike information criterion (AIC) and Bayesian information criterion

(BIC) provide estimates of the relative quality of each fit. Models are ordered from low to high AIC values (best and worst relative fit, respectively). Top right: Model names and their corresponding markers. The colors represent different DNN architectures, the marker shapes represent varying DNN sizes. PyTorch implementations and weights are sourced from TorchVision[41,42]. Classification accuracy is measured on the Imagenet-1k validation set. Straight black lines represent linear fits; the correlation is measured with Spearman's rank correlation $r$ (black text at the top of each figure, corresponding p-values in parentheses).

# Proximity to criticality as an independent predictor of performance

All measures of criticality exhibited strong correlation with DNN performance (Figure 2b). Specifically, while all historical networks operated in the supercritical regime ($\lambda > 0$), closer proximity to the critical state ($\lambda_0 = 0, \Sigma = 0$) was predictive of higher performance. Apart from changes in dynamics, DNNs had also increased in size over time, as measured by the number of modules and parameters ($m$ and $p$, respectively). Consequently, network structure metrics $m$ and $p$ also exhibited correlations with performance, albeit weaker than correlations with dynamics parameters $\lambda_0$, $\Sigma$ (Figure 2c). Because the values of $\lambda_0$ are derived from both the number of modules $m$ and the singular values $\Lambda_0$, we also examined $\Lambda_0$ as dynamical metric independent of structural characteristics, which again predicted performance, indicating a correlation with performance independent from structural metrics (Figure 2b, inset). Figure S1 shows relationships between all other structural and dynamics metrics.

Next, we fitted Linear Mixed Effects Models (LMMs) to further delineate the effects of network dynamics and structure on DNN performance. We evaluated $15$ different LMM configurations - individual models for each predictor $\Sigma, \lambda_0, m, p$ and all possible combinations of these predictors without interaction terms (Figure 2d). The conditional value of determination $R^2$ exceeded $0.87$, indicating that the combination of fixed effects (structural and dynamical measures) and random effects (model type) explained more than $87\%$ of the variance in the network performance. Marginal $R^2$ values, which capture only fixed effects, ranged from $0.07$ to $0.6$, with models incorporating only structural measures (LMM $4, 6, 12$; Figure 2d) again explaining less variance ($7 - 20\%$) compared to models using only criticality measures ($32 - 50\%$, LMM $1, 2, 3$; Figure 2d). The best performing model according to Akaike and Bayesian Information Criterion (AIC and BIC, respectively) incorporated all four predictors, indicating that both structure and dynamics significantly contributed to predicting network accuracy (Figure 2d). The marginal $R^2$ of this best-performing LMM revealed that the combination of structural and dynamical measures alone accounts for more than $60\%$ of the variance in model performance, highlighting the substantial predictive power of these parameters. These results indicated that criticality metrics contributed strongly and independently to prediction performance.

Finally, we demonstrated that optimized training leading to improved performance was associated with closer-to-critical dynamics even when network architectures remained unchanged. For this purpose, we made use of different versions of pre-trained weights belonging to the same model architectures published in the PyTorch Torchvision repository. The original weights $W_0$ closely replicated the results from the original papers introducing each model, whereas the updated version $W_1$ improved these results by using TorchVision's new training recipe introduced in TorchVision v0.11 on October 21, 2021 (recipe comparison in Supplementary Materials). The updated training recipe focused on simplicity and flexibility, such that it could be applied across various architectures, and built upon work in Refs.[43–47]. The improved performances ($\Delta\,\text{accuracy} = \text{accuracy}(W_1) - \text{accuracy}(W_0) > 0$) was again associated with more critical dynamics in most models, evidenced by decreases in both $\Delta\Sigma = \Sigma(W_1) - \Sigma(W_0) < 0$ and $\Delta\Lambda_0 = \Lambda_0(W_1) - \Lambda_0(W_0) < 0$ (Figure 3a, b; details on baseline and improved training performance in Suppl. Table S3). Across models this analysis revealed a statistically significant increase in performance (median 2.81%, interquartile range [IQR] = $1.13\%$, $p = 7.45 \times 10^{-9}$), associated with a significant shift towards more critical dynamics ($\Delta\Sigma$: median = $-9.95 \times 10^{-2}$, IQR = $4.57 \times 10^{-1}$, $p = 2.04 \times 10^{-2}$; $\Delta\Lambda_0$: median = $-2.66 \times 10^{-1}$, IQR = $1.77 \times 10^{1}$, $p = 4.21 \times 10^{-5}$).

Together, these results provide strong evidence that both critical dynamics and structure are crucial factors governing DNN performance, and that state-of-the-art DNN architectures and training inherently promote critical dynamics as an underlying principle driving performance.

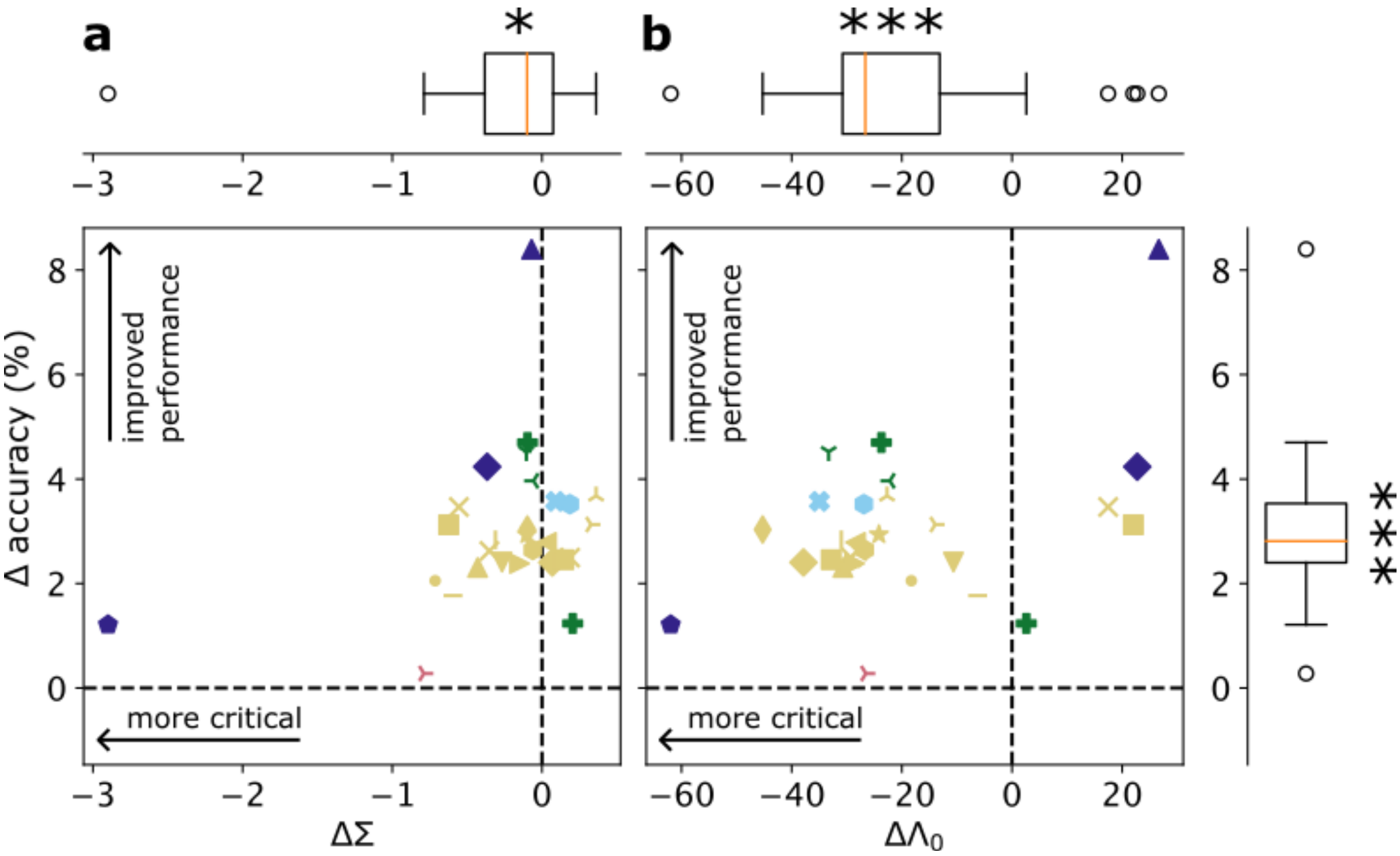

**Figure 3: More critical dynamics is associated with improved DNN performance under different training recipes.** Each marker represents a DNN pretrained on the ImageNet-1K dataset (markers and color codes consistent with those used in Figure 2). $\Delta$ indicates the difference between metrics from original and updated training recipes. Most networks exhibit improved classification performance with updated training recipes ($\Delta\,\text{accuracy} > 0$) alongside more critical dynamics ($\Delta\Sigma < 0$, **a**; $\Delta\Lambda_0 < 0$, **b**). Significance is indicated with asterisks (*** p < 0.001, ** p < 0.01, * p < 0.05).

# Criticality-based training objective improves performance

Building upon the observation that high-performance DNN architectures and training inherently promote critical dynamics, we next developed a novel training approach to explicitly tune neural networks toward criticality by penalizing deviations from critical dynamics in the loss function. Specifically, we evaluated DNN training and performance by comparing standard cross-entropy loss ($L_{CE}$) with a combined criticality-loss $L = L_{CE} + L_\lambda$. Here, $L_\lambda = \alpha(\lambda_t - \lambda_0)^2$ penalized deviations from a target Lyapunov exponent $\lambda_t$ ($\alpha$: scaling factor, $\lambda_0^i$: the Lyapunov exponent in training iteration $i$). We trained each objective using four commonly used initialization schemes: the original initialization from Ref.[48] (init0), the uniform Xavier initialization[13] (init1), the uniform Kaiming initialization[49] (init2), and the initialization from Ref.[5] (Figure 4a, b). With $L_{CE}$ alone, the networks consistently exhibited supercritical dynamics ($\lambda_0 > 0$), as in the example above, while adding the Lyapunov based term tuned the network towards the criticality ($\lambda_0 = 0$, Figure 4a). All test set losses were minimized during training, indicating robust generalization. While $L_{CE}$ replicated the results in Ref.[48], all initialization schemes using the criticality-loss $L$ led to a significant performance increase (Figure 4b, mean over last 10 epochs in inset). The hyperparameters $\alpha$ and $\lambda_t$ controlled both the post-training Lyapunov exponents $\lambda_0$ and the strength of criticality adjustments, as measured by the epoch-wise derivative $\partial L_\lambda / \partial e$ (Figure 4c). This allowed us to tune the networks also to slightly sub- or subcritical dynamics. Optimal performance occurred at criticality ($\lambda_0 = 0$) with modest strength of criticality adjustments. A LMM confirmed that both smaller absolute Lyapunov exponents $|\lambda_0|$ and loss derivates $\frac{\partial L_\lambda}{\partial e}$ significantly improved performance (fixed effect estimates $-0.20, p < 10^{-16}$ and $-0.13, p < 10^{-8}$, respectively). These findings thus demonstrate that optimizing network dynamics through the criticality-loss enhances network performance, which peaks at criticality and yields more robust results that are independent of weight initialization.

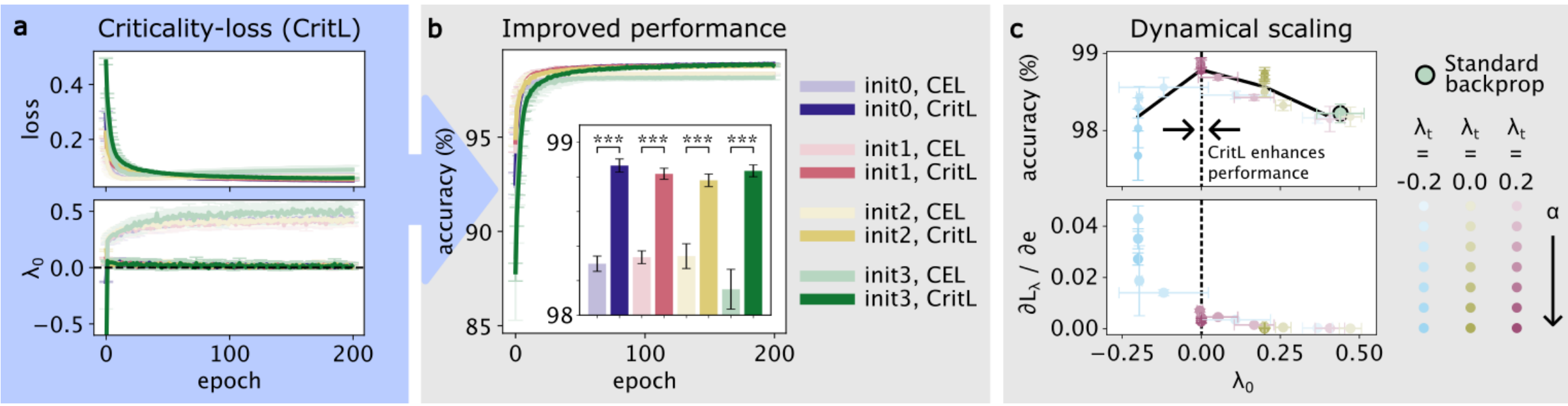

**Figure 4: A novel criticality-loss $L$ (CritL) improves classification performance compared to standard cross-entropy loss $L_{CE}$ (CEL) by tuning DNNs to criticality. a, b,** Evolution of network dynamics during MNIST training across multiple initialization schemes (init0[48], init1[13], init2[49] and init3[50]). **a**, Networks trained with standard cross-entropy loss (light colors) exhibit supercritical dynamics ($\lambda_0 > 0$), while the criticality-loss $L$ (dark colors) achieves criticality ($\lambda_0 \approx 0$) throughout training. **b**, Test set classification accuracy shows enhanced performance of networks trained with criticality-loss $L$. Inset: Average classification accuracies over the final 10 epochs for all runs. Significance is indicated with asterisks (*** p < 0.001). **c**, Parameters $\lambda_t$ and $\alpha$ control dynamics during training, and optimal performance occurs at criticality $\lambda_0 = 0$ (dashed vertical line) with modest criticality-based weight updates measured by $\frac{\partial L_\lambda}{\partial e}$ with training epoch $e$. Parameters: $\lambda_t = [-0.2, 0.0, 0.2]$, $\alpha = [0.01, 0.1, 1, 10, 100, 250, 500]$ trained for 200 epochs. All runs are

averaged across 5 iterations, the solid lines represent the mean and the error bars indicate $\pm 1$ standard error.

# Maintaining criticality prevents performance loss in deep continual learning

Our work so far has demonstrated that DNNs operate predominantly in a slightly supercritical regime and benefit progressively from closer proximity to criticality, which can be enhanced by a dedicated training loss. Despite its successes, deep learning has difficulty adapting to changing data. For instance, large language models are often fine-tuned for specific tasks. Such continual learning on new data, however, reduces the generalizability of these models and leads to reduced performance[18]. The fundamental mechanisms underlying this phenomenon and how to prevent it remain incompletely understood. We now show that loss of performance in deep continual learning is governed by loss of criticality and that maintaining critical dynamics prevents loss of performance.

Following the work by Dohare et al.[18], we used the ImageNet dataset as a testbed for continual learning and constructed a sequence of binary classification tasks by taking classes in pairs (**Figure 5**a). While standard backpropagation rapidly degraded classification accuracy (i.e. loss of plasticity), the continual backpropagation (CBP) algorithm proposed in Ref.[18], which randomly re-initialized a variable small fraction of dormant units (set here by parameter $\rho$), prevented loss of plasticity in a systematic way and dependent on $\rho$ (**Figure 5**b). Importantly, we again observed a systematic relationship between performance and criticality measures. Loss of performance under standard backpropagation was associated with loss of criticality and a strong shift to supercriticality. Conversely, prevention of loss of plasticity using CBP (with increasing $\rho$) moved networks systematically closer to criticality, i.e. closer to $\lambda_0 = 0$, highlighting again a close relationship between DNN accuracy and proximity to the critical point (**Figure 5**b).

The relationship between accuracy and criticality was not simply correlational but causal: systematic control of the distance to criticality using the criticality-loss (CritL) described above again revealed closer proximity to criticality to be associated with higher accuracy, achieving the same accuracy improvements as CBP (**Figure 5**c). Intriguingly, the relationship between accuracy and criticality collapsed on a single curve for CBP and CritL, which indicates that distance to criticality is the common control parameter governing accuracy in both training approaches (**Figure 5**d, top). As in Figure 4c, $\alpha$ and $\lambda_t$ also controlled the strength of criticality adjustments $\partial L_\lambda / \partial e$ (**Figure 5**d, bottom). Task performance is generally determined both by representation learning (controlled by $L_{CE}$) and proximity to criticality (controlled by $L_\lambda$). In line with this and results in Figure 4c, performance in continual learning thus benefitted from closer proximity to criticality when criticality adjustments were moderate (**Figure 5**d, right of dashed line), but dropped slightly when criticality adjustments increased too strongly (**Figure 5**d, left of dashed line). A linear mixed effects model confirmed that both smaller absolute Lyapunov exponents $|\lambda_0|$ and criticality adjustments $\frac{\partial L_\lambda}{\partial e}$ significantly improved performance (fixed effect estimates $-0.24, p < 10^{-8}$ and $-0.50, p < 10^{-16}$, respectively).

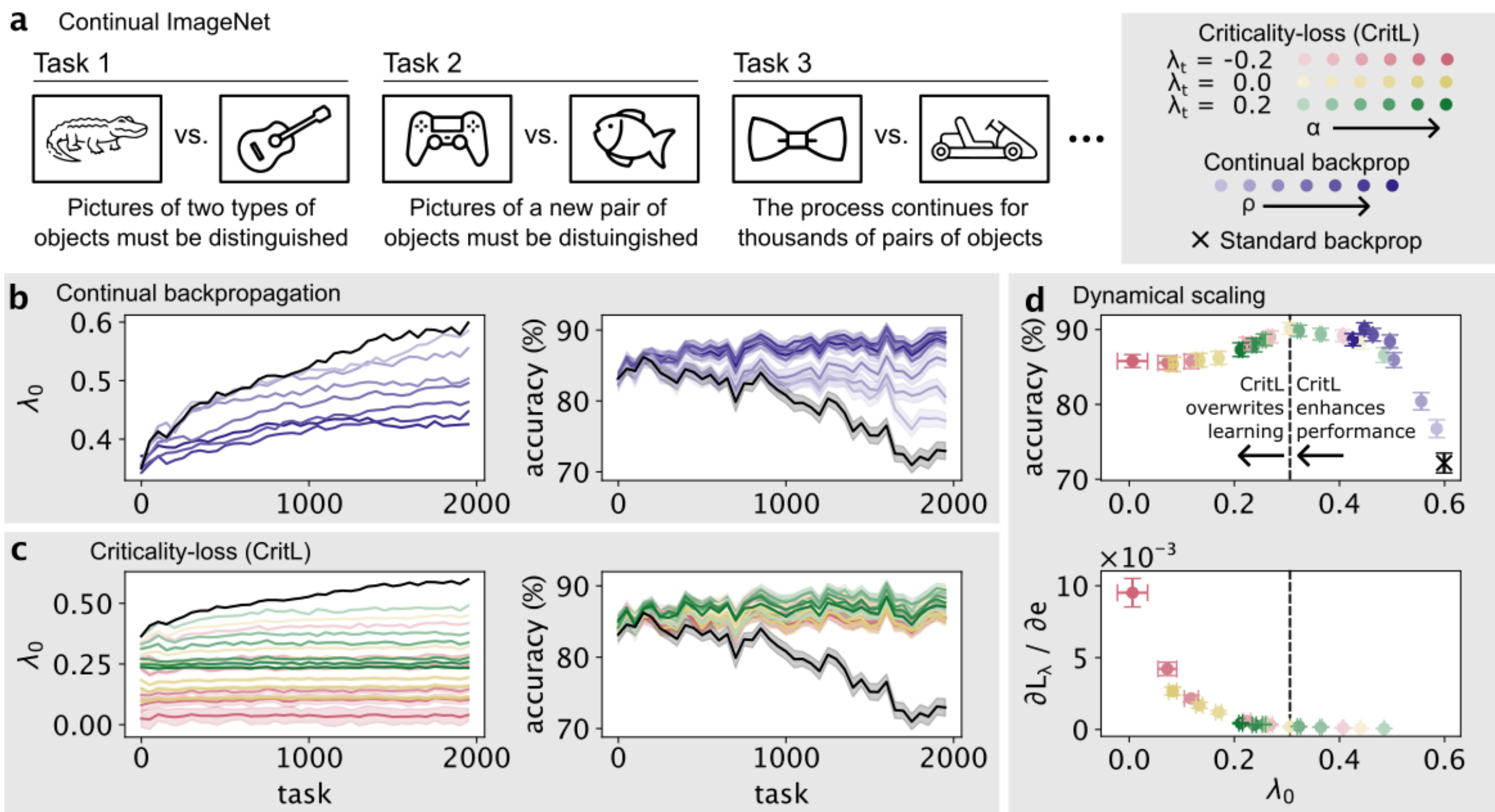

**Figure 5: Maintaining close proximity to criticality prevents loss of performance in continual learning.** **a**, Schematic outline of the continual learning procedure by training a sequence of binary classification tasks using ImageNet pictures (Continual ImageNet). **b**, With standard backpropagation only continual learning leads to gradual loss of plasticity associated with supercritical network dynamics. In contrast, continual backpropagation introduced in Ref[18] preserves plasticity and performance by attenuating the supercritical drift. **c**, Maintaining DNNs close to criticality as the underlying mechanism preventing loss of performance. Directly controlling for critical DNN dynamics by training with criticality-loss (CritL) improves performance and maintains near-critical dynamics (indicated by Lyapunov exponents $\lambda_0 \approx 0$). **d**, Collapse of all training results onto a single performance curve. Parameters $\lambda_t$ and $\alpha$ scale neuron dynamics during training with criticality-loss. Optimal performance (dashed black line) occurs where networks can benefit from proximity to criticality, but where criticality-based weight updates $\partial L_\lambda / \partial e$ are still sufficiently small to not overwrite the learned representations. Parameters: $\lambda_t = [-0.2, 0.0, 0.2]$, $\alpha = [0.01, 0.05, 0.1, 0.5, 1.0, 3.0]$ (CritL), $\rho = [3 \times 10^{-3}, 3 \times 10^{-4}, \ldots, 3 \times 10^{-9}]$ (continual backpropagation), trained for 2,000 tasks, each task for 250 epochs. All results are averaged over 50 tasks; the solid lines represent the mean, shaded regions and error bars correspond to $\pm 1$ standard error. Parts of the figure are adapted from Ref.[18].

# Maintaining criticality prevents model collapse

Next, we show that model collapse, i.e. the declining performance when training on AI-generated content[19], is governed by loss of criticality and that maintaining critical dynamics prevents model collapse.

Following the experimental setup of Shumailov et al.[19], we trained a VAE on MNIST and generated latents from a Gaussian distribution which were then used by the decoder to generate data for subsequent generations. As expected, we observed a progressive decrease of reconstruction quality with generations, where original and generated distributions became increasingly detached from each other, and where the distinct latent clusters disappeared (Figure 6a). We next investigated this process in terms of critical dynamics. In line with our previous observations, we found that networks initially started in a slightly supercritical regime ($\lambda_0 > 0$), but then approached criticality ($\lambda_0 = 0$) around generations $15 - 25$ and abruptly shifted to subcritical dynamics ($\lambda_0 < 0$) after generation $25$ (Figure 6b, dashed vertical line). This transition to subcritical dynamics coincided with a collapse in latent clusters, in output variance $\sigma(pixel) \approx 0$, and in reconstruction accuracy (10%, chance level). These results indicate that model collapse is governed by a non-gradual and sharp state shift that equates a critical transition to subcriticality. Such unwanted catastrophic transitions have already been observed in many other systems ranging from ecosystems to financial markets and the climate[36,37].

Implementing the criticality-loss (CritL) described above during VAE training made the DNN critical, prevented the transition to subcritical dynamics and maintained a meaningful latent clustering and high output variance (Figure 6a, b). Importantly, this approach also led to a maintained high classification accuracy (classification accuracy with CritL 94.7% vs. 10% for vanilla (chance level) after 40 generations; Figure 6b). Parameter scans showed that highest reconstruction quality correlated again with networks being closest to critical dynamics (Figure 6c, d).

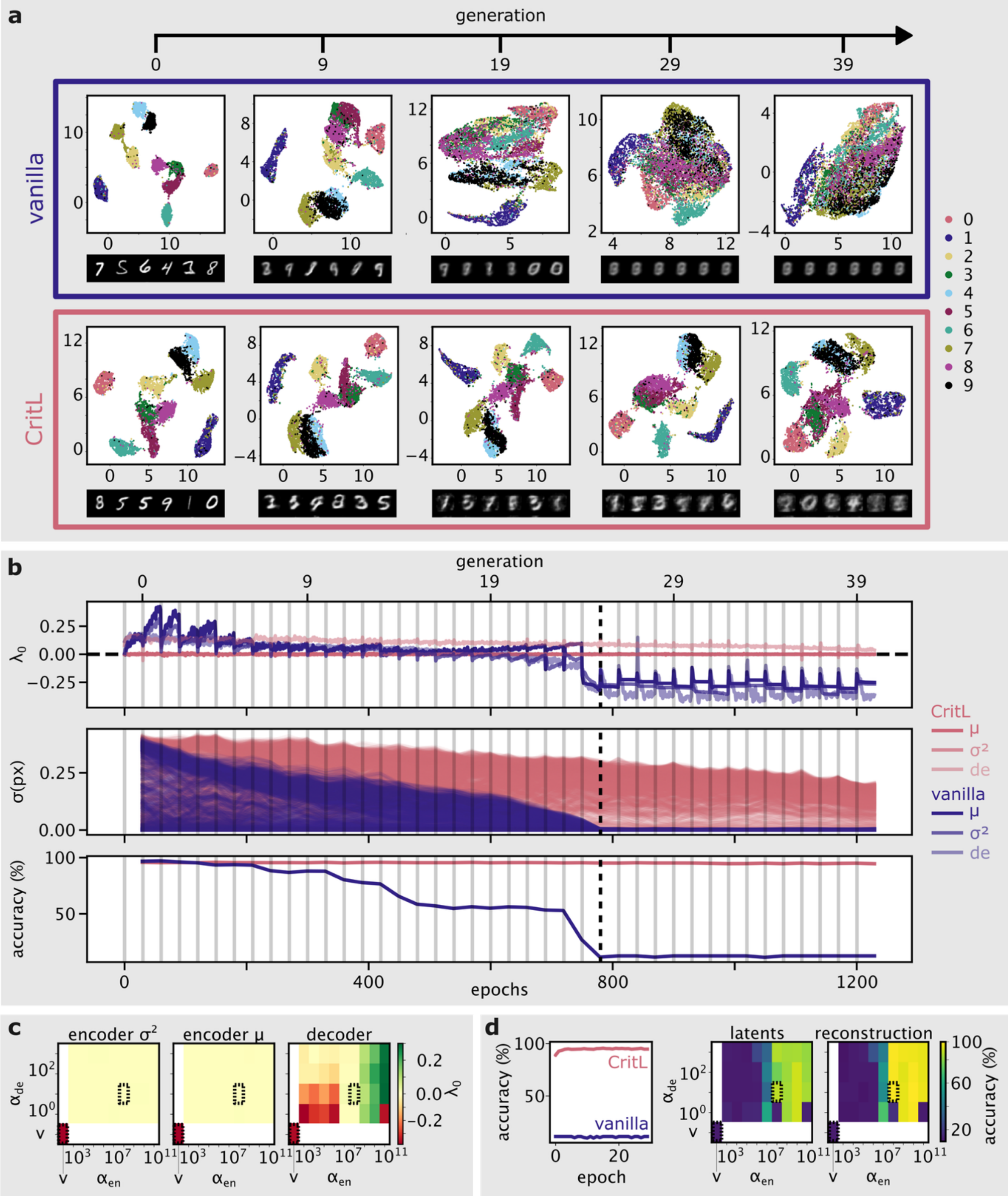

**Figure 6: Critical dynamics prevents model collapse. a,** UMAP visualization of latent space and reconstruction images from MNIST across generations (0, 9, 19, 29, 39) for standard training (blue, vanilla) versus criticality-based loss training (CritL, red). Standard training shows progressive mode entanglement and unimodal collapse (model collapse) after 24 generations, while CritL maintains mode separation and reconstruction quality. **b,** Lyapunov exponents, output variability and reconstruction quality across generations. Standard training (vanilla, blue): All VAE components (encoder heads $\sigma^2$, $\mu$; decoder) approach criticality ($\lambda_0 = 0$, dashed horizontal line) and then abruptly transition to subcritical dynamics, coinciding with collapse in pixel-wise variance $\sigma(px) = 0$ and classification accuracy drop to 10% (chance level) at generation 25 (dashed vertical line). CritL training maintains (near-)critical dynamics and preserves output variability and reconstruction quality. **c,** Parameters scan for CritL scaling factors $\alpha_{en}$ (encoder) and $\alpha_{de}$ (decoder), colored by Lyapunov exponents $\lambda_0$. **d,** Reconstruction quality assessed via classifier accuracy on 39th VAE generation outputs: 30 epochs of classifier training result in 10% (chance level) for standard VAE training (vanilla) and >90% for CritL VAE training. Highest reconstruction quality correlates

with critical dynamics in the parameter scans from panel c. Dashed boxes in the heatmaps indicate runs from panels a, b. Training parameters from VAE and classifier training: 30 epochs each generation, learning rate $= 0.001$, batch size $= 256$, capacity = 64, latent dims = 20.

# Discussion

The rapid advances of modern AI are mostly attributed to three main drivers: larger models, bigger data, more compute. What has been underappreciated is the dynamics of DNNs, contributing to current challenges. The human brain does not seem to suffer these problems. The idea that computation is optimized in the vicinity of phase transitions stems originally from physics[20] and has since had broad impact on the understanding of information processing in biological brain networks[22,23,28]. We here showed that criticality also governs computation in deep artificial intelligence networks, linking network structure to dynamics and function. Looking at DNNs through the lens of a critical phase transition provides a unifying perspective linking the evolution of AI, to understanding heuristically derived algorithm improvements that have implicitly made networks more critical, to engineering novel training objectives that improve robustness and performance by explicitly take criticality into account to preventing model collapse. This perspective provides a new way for quantitatively characterize DNNs and immediately offers solutions to essential AI problems, including loss of plasticity and model collapse, based on a fundamental mechanistic understanding. Our work thus bridges AI with physics and biological cortical network function.

As AI becomes more pervasive, recognizing and reducing the problems associated with synthetic data will be key. The problem of model collapse has remained insufficiently understood with solutions focused mainly on better data curation[51]. Our work establishes model collapse as a catastrophic critical transition to subcriticality[36]. Such unwanted catastrophic transitions have been observed in many other systems; identifying also model collapse as a critical transition thus provides new links to diverse complex systems ranging from ecosystems to financial markets, medicine and the climate governed by these transitions[36,37]. Based on this mechanistic understanding, we show that model collapse can be prevented by keeping DNNs critical, providing the first principled solution to this fundamental AI problem. More generally, this work thus demonstrates that designing effective and performant artificial neural networks requires considering more than just network size: facilitating critical dynamics within the network is essential for enhancing computational performance. While many studies of DNN performance have focused on structural features and dataset properties[2,52–54], our work reveals that network dynamics play an equally crucial role in model performance. By consequence, while the rationale behind many structural and algorithmic improvements has remained vague or intuitive[55–57], we argue that critical phase transitions provide an overarching and theoretically grounded explanation for performance enhancements. Using a large set of historical state-of-the-art object recognition networks demonstrates that optimized training procedures and designs have inherently adopted these principles by driving networks toward criticality. Knowledge of these principles allowed us to explicitly engineer a training method that drives networks to optimal criticality to improve performance and prevent loss of plasticity and model collapse.

Looking forward, criticality metrics like branching parameter $\sigma$ and the Lyapunov exponents, as proposed in this work, may thus help to design better networks or select optimal modules and DNN components by quantifying the distance to the critical point. Utilizing PyTorch's numerical differentiation package these approaches are thus applicable also to much larger networks, including foundation and large language models. Further, as predicted by theory and prior experiments[40], the critical state in DNNs is subject to finite size power law scaling laws, which has significant implications for designing and scaling artificial neural networks. For example, finding the right connectivity at which the critical transition occurs depends on network size and can, in principle, be estimated using these scaling laws. These finite size scaling laws are already used for network optimization[2,53,54]; the criticality framework thus provides a theoretical anchor. Conversely, a deviation from power law scaling has been reported in model collapse[58], which is precisely predicted for a shift to subcriticality.

The observation of optimal dynamics at criticality in DNNs provides a new link between artificial and biological neural systems, as similar relationships between critical dynamics and enhanced functionality have been observed in biological networks[23,28,59]. Strong support for the relevance of a critical branching process governing human brain networks comes from pharmacological studies allowing systematic changes in network connectivity while monitoring its dynamics[27,60]. Insights into how brain networks tune themselves to criticality may help to design similar strategies for DNNs. One such mechanism is sleep, which has been shown as essential to retune networks towards the computationally optimal critical state after a period of wake[29,30]. Further, synaptic plasticity mechanisms, including homeostatic and spike-time-dependent plasticity, have been shown to be able to tune networks to criticality[61]. Future work may therefore explore such local synaptic processes with respect to their ability to self-organize DNNs to criticality. In fact, continual backpropagation (CBP)[18] proposed to prevent loss of plasticity is already very similar to homeostatic plasticity in the sense that both reinitialize inactive units to maintain a homeostatic balance, which in turn, has been shown to lead to global network criticality[40].

In conclusion, our work posits criticality as a unifying framework linking structure, dynamics, and function in deep learning networks, reflecting theoretical predictions and observations in biological systems. This framework provides profound theoretical insights along with immediate practical value (solving major AI challenges). Beyond simply increasing network size, incorporating dynamic optimization towards criticality should become a fundamental design principle for creating more efficient and robust AI. Future research should explore methods to systematically tune and maintain networks at criticality, potentially unlocking new levels of performance and generalization. Integrating these insights from biological neural systems could inspire novel architectures and training protocols that harness the benefits of critical dynamics, advancing the field of deep learning.

# Methods

*Deep neural network models*

We study deep neural networks (DNNs) with respect to critical dynamics and performance. To characterize the phase transition in dynamics, we use simple multi-layer perceptrons with 1,000 layers and constant layer widths each (either 25, 50, 100, 200, 400, 800, or 1,600 neurons per layer; Figure 1). Each neuron is represented by a real-valued, continuous quantity, described by the dynamical equation

$$\mathbf{S}(\mathrm{t}+1) = \mathrm{g}(\mathbf{S}(\mathrm{t}) \cdot \mathbf{W}(\mathbf{t})),$$

where $\mathbf{S}(\mathbf{t}) \in \mathbb{R}^{\mathbf{N}}$ is the neuron state vector at time t, and $\mathbf{W}(\mathbf{t}) \in \mathbb{R}^{\mathbf{N}\times\mathbf{N}}$ the weighted adjacency matrix describing the coupling between the neurons. The activation function is $\mathrm{g}(\mathrm{x}) = \mathrm{a}\tanh(\mathrm{b}\,\mathrm{x})$ with $\mathrm{a} = 1.7159$ and $\mathrm{b} = 0.6666$, which is a configuration often used in the context of deep learning[50,62,63]. The neurons between successive layers are connected by all-to-all and asymmetric synaptic couplings, where the elements $\{\mathrm{w}_{\mathrm{ij}}\} \in \mathbf{W}$ are first drawn from a uniform distribution $[-0.05,0.05]$. To shift the network through different dynamical regimes, the resulting elements $\mathrm{w}_{\mathrm{ij}}$ are then multiplied with a control parameter $\kappa$. All other network architectures (Figures 3-5) are sourced from the literature[18,48,64–83].

*Models pre-trained on the ImageNet-1k database*

ImageNet is an image database organized according to the WordNet hierarchy, where each word is depicted by thousands of images[35]. We investigate neuron dynamics and computational performance of 80 DNNs pre-trained on the ImageNet-1k dataset. ImageNet-1k, also known as the ImageNet Large Scale Visual Recognition Challenge (ILSVRC) 2012, is the most commonly used subset of ImageNet. It comprises 1,281,167 training images, 50,000 validation images, and 100,000 test images across 1,000 object classes. Performance was evaluated based on classification accuracy on the test set. All pre-trained model weights where obtained from the PyTorch Torchvision repository[42]. These pre-trained networks are widely utilized in DNN approaches to facilitate training and serve as foundation models. The networks were implemented using Python PyTorch[41] version 1.13.1 and NumPy[84] version 1.24.2.

*Branching parameter $\sigma$ and finite-time Lyapunov exponents*

To estimate the distance to the critical point we use two measures: the parameter $\sigma$, a measure based on the branching parameter frequently used to characterize branching processes [23,85,86], and the largest finite-time Lyapunov exponent. The calculation of $\sigma$ follows a simple concept: In the subcritical regime, when the connections between neurons are weak, signal activations decrease with each layer and $\sigma$<1. Conversely, in the supercritical regime, strong connections cause signal activations to increase with each layer and $\sigma$>1. At the critical point, activity propagates with $\sigma$=1, avoiding premature die-out or blow-up. To quantify this in DNNs, we define

$$\sigma = \frac{a_{\mathrm{out}}}{a_{\mathrm{in}}},$$

where $a_{\mathrm{in}}$ and $a_{\mathrm{out}}$ represent the average activities of the network's input and output layers, respectively. The average activity of a layer is calculated as

$$a = \frac{1}{N_l}\sum_{i=1}^{N_l} |x_i|,$$

where $N_l$ is the number of neurons in layer $l$, and $x_i$ is the output of neuron $i$. To linearize the relationship between $\sigma$ and classification accuracy, we calculate the distance to criticality ($\sigma = 1$) as

$$\Sigma = |1 - \sigma|.$$

In a dynamical system, the largest Lyapunov exponent describes the time-averaged rate of strongest separation of two infinitesimally close trajectories. In this work, we compute how small input perturbations affect the output of a DNN, exploring an analogy between DNNs and dynamical systems. In this case, the growth or decay of local perturbations is characterized by finite-time Lyapunov exponents. A DNN with $N_0$ input components, $L$ hidden layers, $N_l$ neurons per hidden layer $l = 1, \dots, L$ and $N_{L+1}$ output neurons maps every input $\boldsymbol{x}^{(\mathbf{0})}$ to an output $\boldsymbol{x}^{(\boldsymbol{L}+\mathbf{1})}$. The sensitivity of $x^{(l)}$ to small changes $\delta\boldsymbol{x}$ is determined by the linearization

$$\delta\boldsymbol{x}^{(l)} = \boldsymbol{D}^{(l)}\boldsymbol{W}^{(l)} \dots \boldsymbol{D}^{(2)}\boldsymbol{W}^{(2)}\boldsymbol{D}^{(1)}\boldsymbol{W}^{(1)}\delta\boldsymbol{x} = \boldsymbol{J}_l\delta\boldsymbol{x}.$$

Here, $W^{(l)}$ are the weight matrices in layer $l$, and $D^{(l)}$ are diagonal matrices with elements $D_{jk}^{(l)} = g'\left(b_i^{(l)}\right)\delta_{ij}$, where $b_i^{(l)} = \sum_{j=1}^{N_l} w_{ij}^{(l)} x_j^{(l-1)} - \Theta_i^{(l)}$ and $g'\left(b_i^{(l)}\right) = \frac{d}{db} g(b)\big|_{b=b_i^{(l)}}$ [87]. The function $g(\cdot)$ represents a non-linear activation function, and the weights $w_{ij}$ and thresholds $\Theta_i^{(l)}$ are parameters. The Jacobian matrix $\boldsymbol{J}_l(\boldsymbol{x})$ characterizes the growth or decay of small perturbations to $x$ [88]. Its maximal singular value $\Lambda_0^{(l)}$ increases or decreases exponentially as a function of $l$, with the rate $\lambda_0^{(l)} = l^{-1}\log\left(\Lambda_0^{(l)}(x)\right)$. The singular values $\Lambda_0^{(l)} > \Lambda_1^{(l)} > \dots > \Lambda_{N_l}^{(l)}$ are the square roots of the non-negative eigenvalues of the right Cauchy-Green tensor $\boldsymbol{J}_l^{\top}(\boldsymbol{x})\boldsymbol{J}_l(\boldsymbol{x})$. The maximal eigenvector of $\boldsymbol{J}_l^{\top}(\boldsymbol{x})\boldsymbol{J}_l(\boldsymbol{x})$ determines the direction of maximal stretching, i.e. in which input direction the output changes the most [87]. In practice, we calculate the Jacobian matrix using PyTorch's built-in numerical differentiation package.

*Dynamic range*

Dependent on the stimulus intensity $s$, networks have a minimum response $r_0$ and a maximum response $r_{\mathrm{max}}$. We define the dynamic range $\Delta = 10\,log(s_{0.9}/s_{0.1})$ as the stimulus interval where variations in the stimulus $s$ can be robustly coded by variations in the response $r$, discarding stimuli that are too weak to be distinguished or are too close to saturation [21]. The range of stimuli $[s_{0.1}, s_{0.9}]$ corresponds to the response interval $[r_{0.1}, r_{0.9}]$, where $r_x = r_0 + x\,(r_{\mathrm{max}} - r_0)$, i.e. the stimulation values leading to 10%-90% of response values. To evaluate the dynamic range, we construct an initial neuron state vector $\{s_i\} \in \boldsymbol{S}$, where the entries $s_i$ are drawn from a uniform distribution $[-1, 1]$. Then we multiply all $s_i$ with a scaling parameter, which spans over nine orders of magnitude from $10^{-9}$ to $10^{0}$, corresponding to spontaneous and saturating activity respectively. The network responses are then calculated for each vector $\boldsymbol{s_i}$ as network input.

*Linear mixed-effects models*

We use linear mixed-effects models (LMMs) to evaluate the influence of both dynamical and structural features on classification accuracy. The fixed effects include dynamical parameters (the Lyapunov exponents $\lambda_0$ and the branching parameter $\Sigma$) as well as structural parameters (number of trainable parameters $p$, number of modules $m$). We evaluate models with each fixed effect separately and all possible combinations without interaction terms, to determine the incremental predictive value of adding dynamical parameters for classification accuracy. To account for the variability across the 21 distinct network architecture groups, we incorporate architecture as a random effect, capturing unobserved factors that influence classification accuracy beyond the fixed effects. Model selection was based on Akaike Information Criterion (AIC) and Bayesian Information Criterion (BIC), with lower values indicating better model fit. We also use two LMMs to evaluate the influence of the criticality-based loss (CritL) function on the resulting classification accuracy (standard and continual learning setup). The fixed effects include the maximum Lyapunov exponent $\lambda_0$ and the loss derivate $\partial L_\lambda / \partial e$, a measure of how strong CritL drives the Network towards criticality in each epoch $e$. The random effects comprise the different parameter initialization for $\lambda_t$ and $\alpha$. All LMMs are fitted and evaluated using the lme4 (version 1.1-35.4) [89] and lmeTest (version 3.1-3) [90] packages in R (version 4.4.1) [91]. Models and detailed fit results are available in the Supplementary Materials.

*Model training on the MNIST dataset*

The Modified National Institute of Standards and Technology (MNIST) Dataset served as a benchmark for model evaluation. The dataset comprises 60,000 training images and 10,000 testing images of handwritten digits (0-9), with each grayscale image sized $28 \times 28$ pixel. Pixel intensities of the original images range from 0 (background) to 255 (maximum foreground intensity). The dataset is approximately balanced across all ten-digit classes and has become a standard benchmark for computer vision and machine learning algorithms. We train a simple Multi-Layer Perceptron (MLP) following the approach described in [48], implementing their smallest configuration with 1,000 and 500 neurons in the first and second hidden layer, respectively. Training uses standard backpropagation without momentum and a variable learning rate that shrinks by a multiplicative constant $0.997$ after each epoch, starting from $10^{-3}$. Input preprocessing consisted of mapping pixel intensities to real values in [−1.0, 1.0]. For the activation function we used a scaled hyperbolic tangent $y(a) = A \tanh(Ba)$, where $A = 1.7159$ and $B = 0.6666$. Weights were initialized with four commonly used initialization schemes [13,48,50,64].

*Lyapunov-exponent-based training objective*

We compare two training objectives: First, the standard cross-entropy loss, defined as

$$L_{CE}(x, y) = -\log\left(\frac{\exp(x_y)}{\sum_{c=1}^{C} \exp(x_c)}\right),$$

where $x$ is the network output, $y$ is the target (label), $x_y$ is the logit for the true class and the sum is taken over all classes $C$. Second, we introduced a novel loss term that penalizes deviation from critical phase transitions, defined as

$$L_\lambda(\tilde{x} \to x) = (\lambda_t - \lambda_0(\tilde{x} \to x))^2,$$

where $\tilde{x}$ represents the input image, and $\lambda_0(\tilde{x} \to x)$ denotes the largest finite-time Lyapunov exponent computed for the mapping from input to output. The total loss combines both terms with a scaling parameter $\alpha$ to balance their contributions:

$$L = L_{CE} + \alpha L_{\lambda},$$

where $\alpha$ is a hyperparameter that requires manual tuning. While the cross-entropy loss $L_{CE}$ drives learning the representations, the Lyapunov exponent based term $L_{\lambda}$ guides the networks towards criticality during training. Model performance was evaluated based on classification accuracy on the test set.

*Continual ImageNet*

We investigate neural network performance using Continual Imagenet, a benchmark dataset first introduced in Ref.[18], for evaluating continual learning capabilities in neural networks and their relationship to criticality. The dataset comprises 1,000 classes with 700 images per class, divided into 600 training and 100 test images, enabling systematic evaluation of binary classification tasks. Images are downsampled to $32 \times 32$ resolution to save computation while maintaining sufficient detail for meaningful learning. The dataset facilitates the assessment of various learning algorithms: Standard backpropagation, continual backpropagation and our novel Lyapunov-exponent-based training objective. This enables us to systematically compare different approaches to continual learning with respect to criticality and their effectiveness in mitigating catastrophic forgetting in deep neural networks.

*Continual backpropagation*

Continual backpropagation is a learning algorithm specifically designed to mitigate catastrophic forgetting in deep neural networks first introduced in Ref. [18]. The algorithm extends conventional backpropagation by incorporating selective reinitialization of network units that demonstrate low utility. To quantify unit utility, the algorithm employs a contribution utility measure for each unit in the network. For the $i$-th hidden unit in layer $l$ at time $t$, the contribution utility is defined as

$$\boldsymbol{u}_l[i] = \eta \times \boldsymbol{u}_l[i] + (1 - \eta) \times |\boldsymbol{h}_{l,i,t}| \times \sum_{k=1}^{n_{l+1}} |\boldsymbol{w}_{l,i,k,t}|,$$

where $\eta$ is a decay rate (set to $0.99$), $\boldsymbol{h}_{l,i,t}$ is the output of the $i$-th hidden unit in layer $l$ at time $t$, and $\boldsymbol{w}_{l,i,k,t}$ is the weight connecting the $i$-th unit in layer $l$ to the $k$-th unit in layer $l+1$ at time $t$ and $n_{l+1}$is the number of units in layer $l+1$. This measure estimates a unit's value to its downstream connections by aggregating the product of the unit's activation and its outgoing weights. When a hidden unit's contribution to its consumers is small, the algorithm identifies the unit as non-useful and reinitializes it by setting its outgoing weights to zero. This reinitialization serves two purposes: (1) it neutralizes the unit's influence on the rest of the network, preventing interference with previously learned representations, and (2) it provides the unit with a fresh start, allowing it to potentially learn new or more relevant features. This selective reinitialization mechanism enables the network to maintain important learned features while adaptively repurposing underutilized units for new learning tasks.

*DNN model collapse by training on recursively generated data*

Indiscriminate use of model-generated content in training can lead to degradation of generative performance, a phenomenon referred to as “model collapse”[19]. Following previous work as an example for model collapse, we trained a variational autoencoder (VAE) for 40 generations, each with 30 training epochs. In the first generation, the model was trained on MNIST, in subsequent generations synthetic images generated from random latent vectors of the previous model served as training data. The VAE comprises a convolutional encoder and decoder with a latent dimensionality of 20 and a base channel capacity of 64. Training was performed using the Adam optimizer (learning rate 0.001, batch size 256). The standard VAE loss combines a pixel-wise reconstruction term (binary cross-entropy) with a regularization term that aligns the latent distribution to a unit Gaussian (Kullback–Leibler divergence),

$$L_{VAE} = \underbrace{\sum_{i=1}^{B}\sum_{j=0}^{N} y_{ij} \cdot \log x_{ij} + \left(1 - y_{ij}\right) \cdot \log(1 - x_{ij})}_{binary\ cross-entropy} - \underbrace{\frac{1}{2} \cdot \sum_{i=0}^{B}\sum_{k=0}^{L} 1 + \log \sigma_{ik}^{2} - \mu_{ik}^{2} - \sigma_{ik}^{2}}_{Kullback-Leibler\ divergence},$$

where $B$ is the batch size, $D$ is the number of input features, $x_{ij}$ is the true pixel value of sample $i$ at pixel $j$, $\widehat{x_{ij}}$ the predicted probability, $L$ the dimensionality of the latent space, $\mu_{ik}$ the mean of the posterior distribution $q_{\phi}(z|x^{(i)})$ for sample $i$ in latent dimension $k$, and $\sigma_{ik}^{2}$ the variance of the posterior distribution. We adapt the Lyapunov-exponent-based training objective for VAEs to account for neuron dynamics in both encoder and decoder with

$$L = L_{VAE} + \frac{1}{3}\left[\alpha_{en} \cdot L_{\lambda}(x \to \mu) + \alpha_{en} \cdot L_{\lambda}(x \to \log \sigma^{2}) + \alpha_{de} \cdot L_{\lambda}(x_{l} \to x_{rec})\right],$$

with $L_{\lambda}(x_1 \to x_2) = [\lambda_t - \lambda_0(x_1 \to x_2)^2]$ penalizing the deviation of the maximal Lyapunov exponent from the critical value $\lambda_t = 0$ for the mapping from $x_1$ to $x_2$. Scaling factors $\alpha_{en}$ and $\alpha_{de}$ weight the contributions of the encoder and decoder, respectively. Input images are denoted $x$, latent vectors $x_l$, and reconstructions $x_{rec}$. The two encoder heads output the mean $\mu$ and log-variances $\log \sigma^2$, which parameterize the approximate posterior distribution $q_{\phi}(z|x)$. We evaluate VAE reconstruction quality by training two-layer classifier networks on frozen VAE weights. For latent vectors, we used a small fully connected network (64 and 32 neurons, ReLU activations). For reconstructions, we used a slightly larger fully connected network (100, and 500 neurons, tanh activations). Networks were trained using Adam optimization (learning rate 0.001) with cross-entropy loss. All models were implemented in Python (version 3.10.10) PyTorch (version 2.6.0+cu124) and trained on a GPU.

*Statistical analysis*

We employed multiple statistical tests to evaluate relationships and differences in our measurements. To assess correlation between variables, we used Spearman’s rank correlation coefficient, which makes no assumption about the linearity of relationships and is robust to outliers. For evaluating if observations are distributed around zero, we applied the Wilcoxon signed-rank test, a non-parametric test used to test the location of a population based on a sample of data. For comparing distributions between two groups, we applied the Mann-Whitney U test, a non-parametric test that evaluates whether one distribution is stochastically greater than another. To analyze differences across multiple groups simultaneously, we conducted a one-way Analysis of Variance (ANOVA). Statistical analyses were performed using Python SciPy [92] (version 1.10.1).

# Author contributions

C.M. developed the concept, both authors developed methodology. S.V. conducted simulations, experiments, and data analysis. Both authors wrote the manuscript and contributed to the critical reading and editing.

# Competing interests

The authors declare no competing interests.

# Bibliography

1. LeCun, Y., Bengio, Y. & Hinton, G. Deep learning. *Nature* **521**, 436–444 (2015).
2. Kaplan, J. *et al.* Scaling Laws for Neural Language Models. Preprint at https://doi.org/10.48550/ARXIV.2001.08361 (2020).
3. Hoffmann, J. *et al.* Training Compute-Optimal Large Language Models. Preprint at https://doi.org/10.48550/arXiv.2203.15556 (2022).
4. He, K., Zhang, X., Ren, S. & Sun, J. Deep Residual Learning for Image Recognition. in *2016 IEEE Conference on Computer Vision and Pattern Recognition (CVPR)* 770–778 (IEEE, Las Vegas, NV, USA, 2016). doi:10.1109/CVPR.2016.90.
5. LeCun, Y. *et al.* Backpropagation Applied to Handwritten Zip Code Recognition. *Neural Comput.* **1**, 541–551 (1989).
6. Nair, V. & Hinton, G. E. Rectified Linear Units Improve Restricted Boltzmann Machines. in *International Conference on Machine Learning* (2010).
7. Glorot, X., Bordes, A. & Bengio, Y. Deep Sparse Rectifier Neural Networks. *Proc. Fourteenth Int. Conf. Artif. Intell. Stat.* 315–323 (2011).
8. Ioffe, S. & Szegedy, C. Batch Normalization: Accelerating Deep Network Training by Reducing Internal Covariate Shift. in *Proceedings of the 32nd International Conference on Machine Learning* (eds. Bach, F. & Blei, D.) vol. 37 448–456 (PMLR, Lille, France, 2015).
9. Hochreiter, S. & Schmidhuber, J. Long Short-Term Memory. *Neural Comput.* **9**, 1735–1780 (1997).
10. Hochreiter, S. The Vanishing Gradient Problem During Learning Recurrent Neural Nets and Problem Solutions. *Int. J. Uncertain. Fuzziness Knowl.-Based Syst.* **06**, 107–116 (1998).
11. Bengio, Y. Practical Recommendations for Gradient-Based Training of Deep Architectures. in *Neural Networks: Tricks of the Trade* (eds. Montavon, G., Orr, G. B. & Müller, K.-R.) vol. 7700 437–478 (Springer Berlin Heidelberg, Berlin, Heidelberg, 2012).
12. Introduction. in *Neural Networks: Tricks of the Trade* (eds. Orr, G. B. & Müller, K.-R.) vol. 1524 1–5 (Springer Berlin Heidelberg, Berlin, Heidelberg, 1998).
13. Glorot, X. & Bengio, Y. Understanding the difficulty of training deep feedforward neural networks. *Proc. 13th Int. Conf. Artif. Intell. Stat. AISTATS* **9**,.
14. Ho, A. *et al.* Algorithmic progress in language models. Preprint at https://doi.org/10.48550/arXiv.2403.05812 (2024).
15. Yang, L. & Shami, A. On hyperparameter optimization of machine learning algorithms: Theory and practice. *Neurocomputing* **415**, 295–316 (2020).
16. Feurer, M. & Hutter, F. Hyperparameter Optimization. in *Automated Machine Learning* (eds. Hutter, F., Kotthoff, L. & Vanschoren, J.) 3–33 (Springer International Publishing, Cham, 2019). doi:10.1007/978-3-030-05318-5_1.
17. Narkhede, M. V., Bartakke, P. P. & Sutaone, M. S. A review on weight initialization strategies for neural networks. *Artif. Intell. Rev.* **55**, 291–322 (2022).
18. Dohare, S. *et al.* Loss of plasticity in deep continual learning. *Nature* **632**, 768–774 (2024).
19. Shumailov, I. *et al.* AI models collapse when trained on recursively generated data. *Nature* **631**, 755–759 (2024).
20. Langton, C. G. Computation at the edge of chaos: Phase transistion and

emergent computation. *Phys. Nonlinear Phenom.* **42**, 12–37 (1990).
21. Kinouchi, O. & Copelli, M. Optimal dynamical range of excitable networks at criticality. *Nat. Phys.* **2**, 348–351 (2006).
22. Shew, W. L. & Plenz, D. The Functional Benefits of Criticality in the Cortex. *The Neuroscientist* **19**, 88–100 (2013).
23. Beggs, J. M. & Plenz, D. Neuronal Avalanches in Neocortical Circuits. *J. Neurosci.* **23**, 11167–11177 (2003).
24. Linkenkaer-Hansen, K., Nikouline, V. V., Palva, J. M. & Ilmoniemi, R. J. Long-Range Temporal Correlations and Scaling Behavior in Human Brain Oscillations. *J. Neurosci.* **21**, 1370–1377 (2001).
25. Murray, J. D. *et al.* A hierarchy of intrinsic timescales across primate cortex. *Nat. Neurosci.* **17**, 1661–1663 (2014).
26. Müller, P. M. & Meisel, C. Spatial and temporal correlations in human cortex are inherently linked and predicted by functional hierarchy, vigilance state as well as antiepileptic drug load. *PLOS Comput. Biol.* **19**, e1010919 (2023).
27. Meisel, C. Antiepileptic drugs induce subcritical dynamics in human cortical networks. *Proc. Natl. Acad. Sci.* **117**, 11118–11125 (2020).
28. Müller, P. M., Miron, G., Holtkamp, M. & Meisel, C. Critical dynamics predicts cognitive performance and provides a common framework for heterogeneous mechanisms impacting cognition. *Proc. Natl. Acad. Sci.* **122**, e2417117122 (2025).
29. Meisel, C., Olbrich, E., Shriki, O. & Achermann, P. Fading Signatures of Critical Brain Dynamics during Sustained Wakefulness in Humans. *J. Neurosci.* **33**, 17363–17372 (2013).
30. Meisel, C., Bailey, K., Achermann, P. & Plenz, D. Decline of long-range temporal correlations in the human brain during sustained wakefulness. *Sci. Rep.* **7**, 11825 (2017).
31. Meisel, C. From Neurons to Networks: Critical Slowing Down Governs Information Processing Across Vigilance States. in *The Functional Role of Critical Dynamics in Neural Systems* (eds. Tomen, N., Herrmann, J. M. & Ernst, U.) vol. 11 69–80 (Springer International Publishing, Cham, 2019).
32. Maturana, M. I. *et al.* Critical slowing down as a biomarker for seizure susceptibility. *Nat. Commun.* **11**, 2172 (2020).
33. Meisel, C., Storch, A., Hallmeyer-Elgner, S., Bullmore, E. & Gross, T. Failure of Adaptive Self-Organized Criticality during Epileptic Seizure Attacks. *PLoS Comput. Biol.* **8**, e1002312 (2012).
34. Zimmern, V. Why Brain Criticality Is Clinically Relevant: A Scoping Review. *Front. Neural Circuits* **14**, 54 (2020).
35. Russakovsky, O. *et al.* ImageNet Large Scale Visual Recognition Challenge. *Int. J. Comput. Vis.* **115**, 211–252 (2015).
36. Scheffer, M. *et al.* Early-warning signals for critical transitions. *Nature* **461**, 53–59 (2009).
37. Scheffer, M. *et al.* Anticipating Critical Transitions. *Science* **338**, 344–348 (2012).
38. Harris, T. E. *The Theory of Branching Processes*. (Dover publ, New York, 1989).
39. Zapperi, S., Lauritsen, K. B. & Stanley, H. E. Self-Organized Branching Processes: Mean-Field Theory for Avalanches. *Phys. Rev. Lett.* **75**, 4071–4074 (1995).
40. Bornholdt, S. & Rohlf, T. Topological Evolution of Dynamical Networks: Global Criticality from Local Dynamics. *Phys. Rev. Lett.* **84**, 6114–6117 (2000).

41. Paszke, A. *et al.* PyTorch: An Imperative Style, High-Performance Deep Learning Library. *Adv. Neural Inf. Process. Syst. NeurIPS* (2019).
42. Models and pre-trained weights — Torchvision main documentation. https://pytorch.org/vision/master/models.html.
43. Dollár, P., Singh, M. & Girshick, R. Fast and Accurate Model Scaling. Preprint at http://arxiv.org/abs/2103.06877 (2021).
44. Xiao, T. *et al.* Early Convolutions Help Transformers See Better. Preprint at http://arxiv.org/abs/2106.14881 (2021).
45. Touvron, H., Vedaldi, A., Douze, M. & Jégou, H. Fixing the train-test resolution discrepancy. Preprint at http://arxiv.org/abs/1906.06423 (2022).
46. Touvron, H. *et al.* Training data-efficient image transformers & distillation through attention. Preprint at http://arxiv.org/abs/2012.12877 (2021).
47. Wightman, R., Touvron, H. & Jégou, H. ResNet strikes back: An improved training procedure in timm. Preprint at http://arxiv.org/abs/2110.00476 (2021).
48. Cireşan, D. C., Meier, U., Gambardella, L. M. & Schmidhuber, J. Deep Big Multilayer Perceptrons for Digit Recognition. in *Neural Networks: Tricks of the Trade* (eds. Montavon, G., Orr, G. B. & Müller, K.-R.) vol. 7700 581–598 (Springer Berlin Heidelberg, Berlin, Heidelberg, 2012).
49. He, K., Zhang, X., Ren, S. & Sun, J. Delving Deep into Rectifiers: Surpassing Human-Level Performance on ImageNet Classification. Preprint at http://arxiv.org/abs/1502.01852 (2015).
50. Lecun, Y., Bottou, L., Bengio, Y. & Haffner, P. Gradient-based learning applied to document recognition. *Proc. IEEE* **86**, 2278–2324 (1998).
51. Feng, Y., Dohmatob, E., Yang, P., Charton, F. & Kempe, J. Beyond Model Collapse: Scaling Up with Synthesized Data Requires Verification. Preprint at https://doi.org/10.48550/arXiv.2406.07515 (2024).
52. You, J., Leskovec, J., He, K. & Xie, S. Graph Structure of Neural Networks. Preprint at http://arxiv.org/abs/2007.06559 (2020).
53. Bahri, Y., Dyer, E., Kaplan, J., Lee, J. & Sharma, U. Explaining neural scaling laws. *Proc. Natl. Acad. Sci.* **121**, e2311878121 (2024).
54. Henighan, T. *et al.* Scaling Laws for Autoregressive Generative Modeling. Preprint at https://doi.org/10.48550/ARXIV.2010.14701 (2020).
55. Santurkar, S., Tsipras, D., Ilyas, A. & Madry, A. How Does Batch Normalization Help Optimization? in *Advances in Neural Information Processing Systems* (eds. Bengio, S. et al.) vol. 31 (Curran Associates, Inc., 2018).
56. Shaw, N. P., Jackson, T. & Orchard, J. Biological batch normalisation: How intrinsic plasticity improves learning in deep neural networks. *PLOS ONE* **15**, e0238454 (2020).
57. Van Houdt, G., Mosquera, C. & Nápoles, G. A review on the long short-term memory model. *Artif. Intell. Rev.* **53**, 5929–5955 (2020).
58. Dohmatob, E., Feng, Y., Yang, P., Charton, F. & Kempe, J. A Tale of Tails: Model Collapse as a Change of Scaling Laws. Preprint at https://doi.org/10.48550/arXiv.2402.07043 (2024).
59. Chialvo, D. R. Emergent complex neural dynamics. *Nat. Phys.* **6**, 744–750 (2010).
60. Meisel, C. *et al.* Intrinsic excitability measures track antiepileptic drug action and uncover increasing/decreasing excitability over the wake/sleep cycle. *Proc. Natl. Acad. Sci.* **112**, 14694–14699 (2015).

61. Meisel, C. & Gross, T. Adaptive self-organization in a realistic neural network model. *Phys. Rev. E* **80**, 061917 (2009).
62. Cireşan, D. C., Meier, U., Gambardella, L. M. & Schmidhuber, J. Deep, Big, Simple Neural Nets for Handwritten Digit Recognition. *Neural Comput.* **22**, 3207–3220 (2010).
63. LeCun, Y. Generalization and network design strategies. in (1989).
64. He, K., Zhang, X., Ren, S. & Sun, J. Deep Residual Learning for Image Recognition. (2015) doi:10.48550/ARXIV.1512.03385.
65. Krizhevsky, A., Sutskever, I. & Hinton, G. E. ImageNet Classification with Deep Convolutional Neural Networks. in *Advances in Neural Information Processing Systems* (eds. Pereira, F., Burges, C. J., Bottou, L. & Weinberger, K. Q.) vol. 25 (Curran Associates, Inc., 2012).
66. Liu, Z. *et al.* A ConvNet for the 2020s. Preprint at https://doi.org/10.48550/ARXIV.2201.03545 (2022).
67. Huang, G., Liu, Z., van der Maaten, L. & Weinberger, K. Q. Densely Connected Convolutional Networks. Preprint at https://doi.org/10.48550/ARXIV.1608.06993 (2016).
68. Tan, M. & Le, Q. V. EfficientNet: Rethinking Model Scaling for Convolutional Neural Networks. (2019) doi:10.48550/ARXIV.1905.11946.
69. Tan, M. & Le, Q. V. EfficientNetV2: Smaller Models and Faster Training. (2021) doi:10.48550/ARXIV.2104.00298.
70. Szegedy, C. *et al.* Going Deeper with Convolutions. Preprint at http://arxiv.org/abs/1409.4842 (2014).
71. Szegedy, C., Vanhoucke, V., Ioffe, S., Shlens, J. & Wojna, Z. Rethinking the Inception Architecture for Computer Vision. Preprint at https://doi.org/10.48550/ARXIV.1512.00567 (2015).
72. Tu, Z. *et al.* MaxViT: Multi-Axis Vision Transformer. Preprint at https://doi.org/10.48550/ARXIV.2204.01697 (2022).
73. Tan, M. *et al.* MnasNet: Platform-Aware Neural Architecture Search for Mobile. Preprint at http://arxiv.org/abs/1807.11626 (2019).
74. Sandler, M., Howard, A., Zhu, M., Zhmoginov, A. & Chen, L.-C. MobileNetV2: Inverted Residuals and Linear Bottlenecks. (2018) doi:10.48550/ARXIV.1801.04381.
75. Howard, A. *et al.* Searching for MobileNetV3. Preprint at https://doi.org/10.48550/ARXIV.1905.02244 (2019).
76. Xie, S., Girshick, R., Dollár, P., Tu, Z. & He, K. Aggregated Residual Transformations for Deep Neural Networks. Preprint at http://arxiv.org/abs/1611.05431 (2017).
77. Ma, N., Zhang, X., Zheng, H.-T. & Sun, J. ShuffleNet V2: Practical Guidelines for Efficient CNN Architecture Design. Preprint at http://arxiv.org/abs/1807.11164 (2018).
78. Iandola, F. N. *et al.* SqueezeNet: AlexNet-level accuracy with 50x fewer parameters and <0.5MB model size. Preprint at http://arxiv.org/abs/1602.07360 (2016).
79. Liu, Z. *et al.* Swin Transformer: Hierarchical Vision Transformer using Shifted Windows. Preprint at http://arxiv.org/abs/2103.14030 (2021).
80. Liu, Z. *et al.* Swin Transformer V2: Scaling Up Capacity and Resolution. Preprint at http://arxiv.org/abs/2111.09883 (2022).
81. Simonyan, K. & Zisserman, A. Very Deep Convolutional Networks for Large-Scale Image Recognition. Preprint at http://arxiv.org/abs/1409.1556 (2015).

82. Dosovitskiy, A. *et al.* An Image is Worth 16x16 Words: Transformers for Image Recognition at Scale. Preprint at http://arxiv.org/abs/2010.11929 (2021).
83. Zagoruyko, S. & Komodakis, N. Wide Residual Networks. Preprint at http://arxiv.org/abs/1605.07146 (2017).
84. Harris, C. R. *et al.* Array programming with NumPy. *Nature* **585**, 357–362 (2020).
85. Haldeman, C. & Beggs, J. M. Critical Branching Captures Activity in Living Neural Networks and Maximizes the Number of Metastable States. *Phys. Rev. Lett.* **94**, 058101 (2005).
86. Ribeiro, T. L., Ribeiro, S., Belchior, H., Caixeta, F. & Copelli, M. Undersampled Critical Branching Processes on Small-World and Random Networks Fail to Reproduce the Statistics of Spike Avalanches. *PLoS ONE* **9**, e94992 (2014).
87. Storm, L., Linander, H., Bec, J., Gustavsson, K. & Mehlig, B. Finite-Time Lyapunov Exponents of Deep Neural Networks. *Phys. Rev. Lett.* **132**, 057301 (2024).
88. Strogatz, S. *Nonlinear Dynamics and Chaos: With Applications to Physics, Biology, Chemistry, and Engineering*. (CRC Press, Taylor & Francis Group, Boca Raton, 2024).
89. Bates, D., Mächler, M., Bolker, B. & Walker, S. Fitting Linear Mixed-Effects Models Using lme4. *J. Stat. Softw.* **67**, 1–48 (2015).
90. Kuznetsova, A., Brockhoff, P. B. & Christensen, R. H. B. lmerTest Package: Tests in Linear Mixed Effects Models. *J. Stat. Softw.* **82**, 1–26 (2017).
91. R Core Team. *R: A Language and Environment for Statistical Computing*. (R Foundation for Statistical Computing, Vienna, Austria, 2024).
92. Virtanen, P. *et al.* SciPy 1.0: fundamental algorithms for scientific computing in Python. *Nat. Methods* **17**, 261–272 (2020).